\documentclass[preprint,prb,superscriptaddress]{revtex4}

\usepackage{amsmath}

\usepackage{graphicx,subfigure}

\newcommand{\de}{\epsilon}
\begin{document}

\title{Perturbative and non-perturbative studies with the delta function potential}

\author{Nabakumar Bera}
\affiliation{Department of Theoretical Physics, Indian Association
for the Cultivation of Science, Jadavpur, Kolkata 700 032, India}

\author{Kamal Bhattacharyya}
\affiliation{Department of Chemistry, University of Calcutta, 92
A.P.C. Road, Kolkata 700 009, India}

\author{Jayanta K. Bhattacharjee}
\email{tpjkb@iacs.res.in} \affiliation{Department of Theoretical
Physics, Indian Association for the Cultivation of Science,
Jadavpur, Kolkata 700 032, India}

%\date{\today}

\begin{abstract}
We show that the delta function potential can be exploited along
with perturbation theory to yield the result of certain infinite
series. The idea is that any exactly soluble potential if coupled
with a delta function potential remains exactly soluble. We use the
strength of the delta function as an expansion parameter and express
the second-order energy shift as an infinite sum in perturbation
theory. The analytical solution is used to determine the
second-order energy shift and hence the sum of an infinite series.
By an appropriate choice of the unperturbed system, we can show the
importance of the continuum in the energy shift of bound states.
\end{abstract}

\pacs{03.65.-W, 31.15.Md, 11.55.Hx}

\maketitle

\section{Introduction}\label{intro}
Perturbation theory
\cite{dalgarno,hirschfelder1,wilcox,Reed,Griffith,Bradsten,skala,scherer,hai,killingbeck3}
is a common tool in the study of eigenvalue problems. A classic
eigenvalue problem is the time-independent Schr\"{o}dinger equation,
which can be written as
\begin{equation}\label{0.1}
H\psi=-\frac{\hbar^2}{2m}\nabla^2\psi+V\psi=E\psi,
\end{equation}
where $H$ is the Hamiltonian, $m$ is the mass of the particle, the
wavefunction is $\psi$, and the eigenenergy is $E$. The problem is
to find the allowed energies $E_n$ and the corresponding
wavefunctions $\psi_n$. The energies $E_n$ are generally discrete
and determined by the boundary conditions. Among the very few cases
for which $E$ and $\psi$ can be determined exactly are a particle in
a box, a particle confined by a harmonic potential, and a particle
confined by a Coulomb potential. The shape-invariant potentials for
which exact solutions can be found are variations on these
cases.\cite{khare}

We denote the Hamiltonian for which an exact solution can be found
by $H_0$ and write an arbitrary Hamiltonian $H$ as
\begin{equation}\label{0.2}
H=H_0+\lambda H',
\end{equation}
where $\lambda$ is a parameter and $H'$ is the perturbing
Hamiltonian. We denote the eigenvalues of $H_0$ by ${E_n}^{(0)}$ and
the eigenfunctions by $\psi_n^{(0)}$, and express the eigenvalues
$E_n$ of $H$ in a power series expansion in $\lambda$ as $E_n =
E_n^{(0)}\ + \lambda E_n^{(1)} + \lambda E_n^{(2)} + \ldots$ with
\begin{align}
E_n^{(1)}&= {\langle\psi_n^{(0)}|H'|\psi_n^{(0)}\rangle},
\label{0.4}\\
E_n^{(2)} &=
\sum_{l\neq{}n}\frac{|\langle\psi_n^{(0)}|H'|\psi_l^{(0)}\rangle|^2}{E_n^{(0)}
- E_l^{(0)}}.\label{0.5}
\end{align}
Equations~\eqref{0.4} and \eqref{0.5} are valid when $\psi_n^{(0)}$
is nondegenerate. From Eq.~(\ref{0.5}) we see that the expression
for $E_n^{(2)}$ is an infinite sum. If we choose the perturbation
such that both $H$ and $H_0$ are exactly soluble, we can obtain an
exact expression for $E_n^{(2)}$ which, when equated with the
expression in Eq.~(\ref{0.5}), leads to an infinite series whose sum
is known.\cite{simon}

The $\delta$-function potential can be used to fulfill this purpose.
The solutions of $H\psi=E\psi$ are the same as the solutions of
$H_0\psi=E_0\psi$ except where the $\delta$-function is nonzero, and
$E$ can be found by matching the solutions across the
$\delta$-function. This procedure can provide some other insights as
well. In Sec.~II we discuss an example where the continuum states
can change the sign of the contribution coming from the bound
states. Another example is the effect of physical confinement on
bound structures such as molecules. In this case we can study the
effect of enclosing the $\delta$-function(s) between two infinite
walls which is explored in Problem 1 in Sec. VI. A similar extension
of Problem 3, given in Sec. VI is also possible.

In Sec.~II we consider a box with infinite walls perturbed by an
attractive $\delta$-function potential placed anywhere within the
box. Next we consider a box of finite depth so that the number of
bound states is finite. In this case the impact of the continuum
states on the calculation of $E_n^{(2)}$ becomes of primary
interest. In Sec.~III we use the simple harmonic oscillator with a
$\delta$-function potential at the origin to obtain an interesting
infinite series and show how we can do the same for the
one-dimensional hydrogen atom. The convergence of some other series
is briefly considered in Sec.~IV. We give a few suggested problems
in Sec.~VI.

\section{Particle in a box with delta function(s)}\label{II}

The $\delta$-function potential has been used as a model for atoms,
molecules, solids, liquids, the many body problem, and
scattering.\cite{lapidus1} The usefulness of the $\delta$-function
potential has been discussed by Lapidus\cite{lapidus2} in the
context of stationary state perturbation theory. We place the
potential at an arbitrary position, $x=pL$ with $0<p<1$, in a box of
length $L$, with $0\leq x \leq L$. The general solution of
Eq.~\eqref{0.1} can be written as
\begin{equation}
\psi(x) =
\begin{cases}
A\sin kx +B\cos kx, & (0\leq x \leq pL) \\
C\sin kx +D\cos kx, & (pL\leq x\leq L),
\end{cases}
\end{equation}
where $k=\sqrt{2mE/\hbar^2}$ and $A$, $B$, $C$, $D$ are determined
from the normalization and boundary conditions. Because $\psi(x)=0$
at $x=0$, we have $B=0$. Hence we can write
\begin{equation}\label{1.4}
\psi(x)=A\sin kx, \qquad (0\leq x \leq pL).
\end{equation}
We also have $\psi(x=L)=0$ so that $D=-C\tan kL$ and
\begin{equation}\label{1.5}
\psi(x)=\tilde{C}\sin k(L-x), \qquad (pL\leq x \leq L)
\end{equation}
where $\tilde{C}=C/\cos(kL)$. The continuity condition of the two
functions at $x=pL$ gives
\begin{equation}\label{1.6}
\frac{A}{\tilde{C}}=\frac{\sin k(1-p)L}{\sin kpL}.
\end{equation}

If we integrate Eq.~(\ref{0.1}) from $-\epsilon$ to $+\epsilon$ and
take the limit as $\epsilon\rightarrow 0$
\begin{equation}
\frac{d\psi}{dx}|_{+\epsilon}-\frac{d\psi}{dx}|_{-\epsilon}=
\frac{2m}{\hbar^2}\lim_{\epsilon\rightarrow
0}\int_{-\epsilon}^{+\epsilon}V(x)\psi(x).
\end{equation}

It is clear that if $V(x)$ is finite, $\frac{d\psi}{dx}$ is
continuous at $x=0$. But when $V(x)$ is infinite this agreement
fails. In particular if $V(x)=-\lambda\delta(x)$

\begin{equation}
\frac{d\psi}{dx}|_{+\epsilon}-\frac{d\psi}{dx}|_{-\epsilon}=
-\frac{2m\lambda}{\hbar^2}\psi(0).
\end{equation}

Using this discontinuity condition at $x=pL$ and Eq.~(\ref{1.6}), we
obtain the eigenvalue condition,
\begin{equation}\label{1.7}
k\sin kL=\frac{2m\lambda}{\hbar^2}\sin kpL \sin k(1-p)L.
\end{equation}
If the perturbation is small, $k$ can be expanded as
\begin{equation}\label{1.8}
k=k^{(0)} + \lambda k^{(1)} + \lambda^2k^{(2)} +\ldots
\end{equation}
We write
\begin{equation}
\label{1.9} E_n =\frac{\hbar^2}{2m}k^2 =\frac{\hbar^2}{2m}
\left[k^{(0)} + \lambda k^{(1)} + \lambda^2k^{(2)} +
\ldots\right]^2,
\end{equation}
which gives
\begin{subequations}
\label{1.10}
\begin{align}
E_n^{(0)}&=\frac{\hbar^2}{2m}{k^{(0)}}^2,\\
E_n^{(1)}&=\frac{\hbar^2}{2m}2k^{(0)}k^{(1)},\\
E_n^{(2)}&=\frac{\hbar^2}{2m}\left[{k^{(1
)}}^2+2k^{(0)}k^{(2)}\right].
\end{align}
\end{subequations}
From Eqs.~(\ref{1.7}) and (\ref{1.8}) we obtain
\begin{subequations}
\label{1.11}
\begin{align}
k^{(0)}&=\frac{n\pi}{L},\\
k^{(1)}&=(-1)^n\frac{2m}{k^{(0)}L\hbar^2}\sin^2(k^{(0)pL}),\\
k^{(2)}&=-\frac{k^{(1)}}{k^{(0)}}+\frac{2mk^{(1)}}{k^{(0)}\hbar^2}
\left[\sin(k^{(0)}pL)\cos(k^{(0)}pL)(1-2p)\right],
\end{align}
\end{subequations}
so that $E_n^{(0)}=n^2\hbar^2\pi^2/2mL^2$ and
\begin{subequations}
\begin{align}
E_n^{(1)}&=-\frac{2}{L}\sin^2(n\pi p), \label{1.12}\\
E_n^{(2)}&=-\frac{2m}{n^2\hbar^2\pi^2}\sin^4(n\pi
p)\left[1+2\pi{n}(1-2p)\cot(n\pi{p})\right]\label{1.13}.
\end{align}
\end{subequations}

If we substitute the energy eigenvalues and eigenfunctions of the
unperturbed Hamiltonian and write the perturbation as
$H'=-\lambda\delta(x-pL)$ in Eq.~(\ref{0.5}), we find
\begin{equation}\label{1.15}
E_n^{(2)}=\frac{8m}{\hbar^2\pi^2}\sin^2(n\pi p)\sum_{l\neq
n}\frac{\sin^2(l\pi p)}{n^2-l^2}.
\end{equation}
From Eqs.~(\ref{1.13}) and (\ref{1.15}), we find
\begin{equation}\label{1.16}
4\sum_{l\neq{n}}\frac{\sin^2(l\pi p)}{l^2-n^2}= \sin^2(n\pi
p)\left[\frac{1}{n^2}+2\frac{\pi(1-2p)}{n}\cot(n\pi p)\right].
\end{equation}
If $l$ and $n$ are odd integers and $p=1/2$, Eq.~\eqref{1.16}
simplifies to
\begin{equation}\label{1.17}
4\sum_{l\neq n }\frac{1}{l^2-n^2}= \frac{1}{n^2}\qquad \mbox{($l$,
$n$ odd)}.
\end{equation}

Equation~\eqref{1.17} is identical to Eq. $(A_3)$ of
Ref.~\onlinecite{lapidus2}. To verify that the sum is correct and
the series is convergent, we write Eq.~\eqref{1.17} as
\begin{subequations}
\label{1.17e}
\begin{align}
4\sum_{l\neq n}\frac{1}{l^2-n^2}&=
\frac{2}{n}\lim_{N\rightarrow\infty}\Big[\sum_{m\neq\frac{n-1}{2}}^N\frac{1}
{(2m+1)-n}-\sum_{m\neq \frac{n-1}{2}}^N\frac{1}{(2m+1)+n} \Big]\\
&= \frac{1}{n^2}+\lim_{N\rightarrow\infty}
\Big[\sum_{{\frac{1-n}{2}}}^{-1}\frac{1}
{p}+\sum_{-1}^{{\frac{n-1}{2}}}\frac{1}
{p}+\sum_{{\frac{n+1}{2}}}^{N-\frac{n+1}{2}}\frac{1}{p}-
\sum_{{\frac{n+1}{2}}}^{N+\frac{n+1}{2}}\frac{1}{q} \Big]\\
&= \frac{1}{n^2}+\lim_{N\rightarrow\infty}
\Big[\sum_{{\frac{n+1}{2}}}^{N-\frac{n+1}{2}}\frac{1}{p}-
\sum_{{\frac{n+1}{2}}}^{N+\frac{n+1}{2}}\frac{1}{q} \Big]
=\frac{1}{n^2},
\end{align}
\end{subequations}
because for finite $n$, $N-(n-1)/2$ and $N+(n+1)/2$ both approach
$N$ as $N\rightarrow \infty$.

We can employ the first-order correction term to the wavefunction,
which also has a sum-over-states form, to obtain a sum rule. We have
\begin{equation}
\label{1.18} \psi(x)=
\begin{cases}
A\sin kx, & (0\leq x \leq pL) \\
A\frac{\sin kpL}{\sin k(1-p)L}\sin k(L-x) & (pL\leq x \leq L)
\end{cases}
\end{equation}
with
\begin{equation}\label{1.19}
A^2\left[\frac{L}{2}-\frac{\sin kL\cos2kL(1-2p)}{2k}\right]=1
\end{equation}
from the normalization condition. If we write $k$ as in
Eq.~(\ref{1.8}) and write $\psi =\psi^{(0)} + \lambda \psi{(1)} +
\ldots$, the first-order shift in the wavefunction can be found from
\begin{align}\label{1.20}
\psi(x)& =
\frac{\sin{k^{(0)}{x}}+{\lambda}k^{(1)}{x}\cos{k^{(0)}x}}{\sqrt{\frac{L}{2}}
\left[1-\frac{{\lambda}k^{(1)}{L}}{k^{(0)}{L}}\cos{k^{(0)}{L}}\cos{2k^{(0)}{L}(1-2p)}\right]^{1/2}} \\
& \approx \sqrt{\frac{2}{L}}\sin
k^{(0)}{x}\left[1+\frac{{\lambda}k^{(1)}}{2k^{(0)}}(-1)^n\cos{4np\pi}\right]
+\lambda\sqrt{\frac{2}{L}}k^{(1)}{x}\cos{k^{(0)}{x}}+O(\lambda^{2})
\end{align}
Thus the first-order shift $\psi_n^{(1)}(x)$ is
\begin{equation}\label{1.21}
\psi_n^{(1)}(x)=\sqrt{\frac{2}{L}}\sin{\frac{n\pi{x}}{L}}\frac{k^{(1)}}{2k^{(0)}}(-1)^n\cos
4np\pi +\sqrt{\frac{2}{L}}k^{(1)}{x}\cos{\frac{n\pi{x}}{L}}.
\end{equation}
If we use the fact that $H'_{ln}=-\frac{2}{L}\sin n p \pi\sin l
p\pi$, where $H'_{ln}=\langle \psi_l^{(0)}|H|\psi_n^{(0)}\rangle$,
we can write Eq.~(\ref{0.4}) as
\begin{equation}\label{1.22}
\psi_n^{(1)}(x)=-\sqrt{\frac{2}{L}}\frac{2}{L}\sum_{l\neq
n}\frac{\sin n p\pi \sin l p\pi
\sin{\frac{l\pi{x}}{L}}}{E^{(0)}_n-E^{(0)}_l},
\end{equation}
which leads to
\begin{eqnarray}\label{1.23}
-\sqrt{\frac{2}{L}}\frac{2}{L}\sum_{l\neq n}\frac{\sin n p\pi \sin l
p\pi \sin{\frac{l\pi{x}}{L}}}{\frac{\pi^2\hbar^2}{2mL^2}(n^2-l^2)}
& =&\sqrt{\frac{2}{L}}\sin{\frac{n\pi{x}}{L}}\frac{k^{(1)}}{2k^{(0)}}(-1)^n\cos{4np\pi} \nonumber \\
&&{}+\sqrt{\frac{2}{L}}k^{(1)}{x}\cos{\frac{n\pi{x}}{L}},
\end{eqnarray}
with $l$ and $n$ odd and $0\leq x\leq pL$.

The effect of continuum states in second-order perturbation theory
has been considered in Refs.~\onlinecite{schiff, landau, kiang,
coronado, urumov}. A one-dimensional attractive $\delta$-function
potential was considered in Ref.~\onlinecite{kiang}. This potential
supports only one bound state\cite{Merzbacher} and for a given
perturbation, the contribution to $E_n^{(2)}$ comes only from the
continuum states. A one-dimensional square well potential perturbed
by another square well potential was discussed in
Ref.~\onlinecite{coronado}.

In this context we discuss an example where the result of
$E_n^{(2)}$ is a sum of a finite series for a finite number of bound
states and an integral from the continuum. The continuum
contribution is so large that it reverses the sign of the sum. We
consider a finite square well potential, $V=0$ for $-L<x<L$ and
$V=V_0$ outside for which there is a finite number of bound states
determined by the magnitude of $V_0$.\cite{Griffith, Merzbacher,
griffith1} In the following we will consider the bound state with
the highest energy. The perturbation is an attractive
$\delta$-function at $x=0$. Because all the other bound states are
lower in energy than the one we are considering, their contribution
to the sum in Eq.~(\ref{0.5}) is positive.

Equation~\eqref{0.1} outside the well can be written as
\begin{equation}
\frac{d^2}{dx^2}\psi(x)+ \kappa^2\psi(x)=0,
\end{equation}
with
\begin{equation}
\kappa^2=\frac{2m}{\hbar^2}(V_0-E).
\end{equation}
 The corresponding solution is
\begin{equation}
\psi(x)=
\begin{cases}
B\exp(-\kappa x) & (x>0),\\
B\exp(\kappa x) & (x<0).
\end{cases}
\end{equation}
Equation~\eqref{0.1} inside the well can be written as
\begin{equation}
\frac{d^2}{dx^2}\psi(x)+{k}^2\psi(x)=0,
\end{equation}
where
\begin{equation}
{k}^2=\frac{2mE}{\hbar^2}.
\end{equation}
The corresponding solutions are
\begin{equation}
\psi(x)=
\begin{cases}
A\cos kx, & \mbox{even} \\
A\sin kx, & \mbox{odd}.
\end{cases}
\end{equation}
The acceptability conditions for the wavefunctions are equivalent to
requiring that the logarithmic derivative of the wavefunction be
continuous at $x=\pm L$. For even states this condition leads to
\begin{subequations}
\begin{align}
\tan kL & =\frac{\kappa}{k}=\sqrt{\frac{\alpha^2L^2}{k^2L^2}-1}\\
\alpha&=\sqrt{\frac{2mV_0}{\hbar^2}}.
\end{align}
\end{subequations}
If we plot $\alpha^2L^2-k^2L^2$ and $\sqrt{k^2L^2}\tan^2
\sqrt{k^2L^2}$ versus $k^2L^2$, the intersections will give the
scaled values of the bound state energies. Here all bound states are
non-degenerate. Even and odd solutions alternate as the energy
increases. The number of bound states is finite and equal to $N+1$
if $N\pi<2\alpha L \leq(N+1)\pi$. Hence, by selecting a value of
$\alpha$, that is, $V_0$, we can control the number of bound states.

If we perturb the system by placing an attractive $\delta$-function
potential at $x=0$, the wavefunction is given as
\begin{equation}
\psi(x)=
\begin{cases}
A\cos(kx)+B\sin kx & (0< x \leq L) \label{41}\\
A\cos(kx)-B\sin kx & (0>x\geq -L)\\
C\exp(-\kappa|x|) \quad & (x>|L|)
\end{cases}
\end{equation}
The continuity condition at $x=0$ gives $B=m\lambda A/k\hbar^2$, and
the wavefunction inside the well is
\begin{equation}
\psi(x)=A\big[\cos kx+\frac{m\lambda}{k\hbar^2}\sin kx\big].
\end{equation}
The continuity condition at $x=+L$ gives
\begin{equation}
k^2\tan kL -k \kappa =\lambda\frac{m}{\hbar^2}[\kappa\tan kL +k].
\end{equation}
If we substitute the values of $k$ and $\kappa$ given in Eqs. (32)
and (29) in Eq. (37), we obtain the desired result
\begin{equation}
E\tan \Big(\sqrt{2mE}\frac{L}{\hbar}
\Big)-\sqrt{E(V_0-E)}=\lambda\frac{\sqrt{m}} {\sqrt 2 \hbar}
\Big[\sqrt{V_0-E}\tan(\sqrt{2mE}\frac{L}{\hbar})+\sqrt E \Big].
\end{equation}
We expand $E$ in powers of $\lambda$, equate terms of same order of
$\lambda$ on both sides and obtain
\begin{subequations}
\begin{align}
\tan\Big(\sqrt{2mE_n^{(0)}}\frac{L}{\hbar} \Big)&=\sqrt{\frac{\de}{E_n^{(0)}}},\\
E_n^{(1)}&=\frac{\sqrt{\frac{2m\de}{\hbar}}}
{1+L{\sqrt{\frac{2m \de}{\hbar}}}},\\
E_n^{(2)}&=-\frac{V_0}{{E_n^{(0)}}^{3/2}}
\frac{1}{\frac{LV_0\sqrt{m}}{\hbar\sqrt{2E_n^{(0)}}}+
\frac{1}{2}\sqrt{\frac{\de}{E_n^{(0)}}}+\frac{1}{2}
\sqrt{\frac{E_n^{(0)}}{\de}}}\nonumber\\
& \times \frac{\frac{1}{2}m^{3/2}L \de +\frac{m^2L^2 \de^{3/2}}
{\sqrt{2}\hbar} +\frac{m\hbar \de}{2\sqrt{2}}(2E_n^{(0)}-V_0)+
\frac{m^{3/2}L \de}{2}(2E_n^{(0)}-V_0)}{\sqrt{2}
\hbar^2\big[\hbar+L\sqrt{2m \de}\big]\big[1+L^2\frac{2m\de}{\hbar^2}
+\frac{2L}{\hbar}\sqrt{2m \de}\big]}, \label{100}
\end{align}
\end{subequations}
where $\de = V_0 - E_n^{(0)}$. For the $n$th state to be bound
$E_n^{(0)}$ must be less than $V_0$. From the solution for
$E_n^{(0)}$  from Eq.~(39a) we see that
$E_n^{(0)}\leq\frac{1}{2}{V_0}$, even for the most excited bound
state. Given these constraints, we conclude that $E_n^{(2)} < 0$,
which makes it clear that the integral from the continuum states has
a greater contribution than the discrete sum. If we fix $V_0$ and
$L$ such that there are only three bound states with energies $E_1$,
$E_2$, and $E_3$, then wavefunctions corresponding to $E_1$ and
$E_3$ are even and that corresponding to $E_2$ is odd. The
contribution from the bound states to the energy shift of $E_3$ to
second order is
\begin{equation}
E_3^{(2)}=\frac{A_1A_3}{E_3-E_1},
\end{equation}
where $A_1$ and $A_3$ are defined in Eq.~(\ref{41}) for the ground
state and the second excited state respectively with a similar
meaning for $k_1$ and $k_3$. The relation between $A$ and $k$ is
given by
\begin{equation}
A_{1,3}=\frac{\sqrt{2}}{\sqrt{L\pm\frac{1}{2k}\sin(2k_{1,3}L)+\cos^2(\frac{k_{1,3}L}{2\kappa})}},
\end{equation}
which is positive, where $+$ and $-$ signs are for $A_1$ and $A_3$
respectively. The contribution from the continuum makes the final
answer as given by Eq.~(\ref{100}) and is negative, showing the
important role of the continuum.

\section{Particle confined by a harmonic potential}\label{V}

We now consider the exactly soluble system of a simple harmonic
oscillator perturbed by an attractive delta potential at the origin.
The unperturbed Hamiltonian is $H_0=p^2/2m+ kx^2/2$, and the
perturbation is $H'=-\lambda\delta(x)$. Equation~\eqref{0.1} for the
unperturbed system can be solved as
\begin{equation}\label{3.2}
\frac{d^2}{dz^2}\psi(z)+\left(n +
\frac{1}{2}-\frac{z^2}{4}\right)\psi(z)=0,
\end{equation}
where $z=\sqrt{2m\omega/\hbar}x$ and $E_n=(n+1/2)\hbar\omega$,
subject to the boundary condition that the wavefunction vanishes at
$\pm\infty$. The even and odd solutions of the unperturbed system
are
\begin{subequations}
\begin{eqnarray}
y_1 =e^{-z^2/4}
{_1}F_1\left(\frac{a}{2}+\frac{1}{4};\frac{1}{2};\frac{z^2}{2}\right)
\mbox{even},\\
y_2=ze^{-z^2/4}{_1}F_1\left(\frac{a}{2}+\frac{3}{4};\frac{3}{2};\frac{z^2}{2}\right)
\mbox{odd},
\end{eqnarray}
\end{subequations}
where the true solution $\psi$ is a linear combination of $y_1$ and
$y_2$ (see Eq. (47)) and $_1F_1$ is the confluent hypergeometric
function (the Kummer function). Its power series expansion can be
written as
\begin{equation}
_1F_1\left(a;b;z\right)=1+\frac{a}{b}\frac{z}{1!}+\frac{a(a+1)}{b(b+1)}\frac{z^2}{2!}
+ \ldots
\end{equation}
The solution can be written in the usual form as
\begin{subequations}
\label{3.11}
\begin{align}
U(a,z)&=\cos\pi\Big(\frac{1}{4}+\frac{a}{2}\Big)Y_1 -
\sin\pi\Big(\frac{1}{4}+\frac{a}{2}\Big)Y_2,\\
V(a,z)&=\frac{1}{\Gamma\big(\frac{1}{2}-a \big)} \Big[\sin\pi
\Big(\frac{1}{4}+\frac{a}{2} \Big)Y_1 - \cos\pi
\Big(\frac{1}{4}+\frac{a}{2})Y_2 \Big],
\end{align}
\end{subequations}
where $Y_1$ and $Y_2$ are given by
\begin{subequations}
\begin{align}
Y_1&=\frac{1}{\sqrt{\pi}}\frac{\Gamma(\frac{1}{4}-\frac{a}{2})}{2^{(\frac{a}{2}+\frac{1}{4})}}
y_1,\\Y_2&=\frac{1}{\sqrt{\pi}}\frac{\Gamma(\frac{3}{4}-\frac{a}{2})}
{2^{(\frac{a}{2}+\frac{1}{4})}}y_2 .
\end{align}
\end{subequations}

We write the general solution for the perturbed problem as
\begin{equation}\label{3.17}
\psi =
\begin{cases}
Ay_1+By_2 & (x>0) \\
Cy_1+Dy_2 & (x<0).
\end{cases}
\end{equation}
 These two forms of $\psi$ must match at $x=0$, giving
$A=C$. The wavefunction must be either symmetric or antisymmetric.
The antisymmetric functions remain unchanged even after introducing
the delta potential. The symmetric functions must change and are
expressed as
\begin{equation}\label{3.18}
\psi =
\begin{cases}
Ay_1+By_2,\\
Ay_1-By_2.
\end{cases}
\end{equation}
The discontinuity condition of the $\delta$-function at the origin
gives $B=-m\lambda\xi A/\hbar^{2}$, where
$\xi=\sqrt{\hbar/2m\omega}$, and the acceptable wavefunction takes
the form
\begin{equation}\label{3.19}
\psi = A \Big[y_1-\frac{m\lambda \xi}{\hbar^{2}}y_2 \Big].
\end{equation}
The boundary condition $\psi\rightarrow 0$ as $x\rightarrow\infty$
leads to
\begin{equation}\label{3.20}
y_1-\frac{m\lambda \xi}{\hbar^{2}}y_2=0.
\end{equation}
If we compare this form with Eq.~(45a), we find that
$U(a,x)\rightarrow 0$ as $x\rightarrow\infty$ and thus  we obtain
\begin{equation}\label{3.20}
\sqrt{2}\frac{\Gamma(\frac{3}{4}-\frac{a}{2})}{\Gamma(\frac{1}{4}-\frac{a}{2})}
\tan{(\frac{\pi}{4}+\frac{{\pi}a}{2})}=-\frac{m\lambda
\xi}{\hbar^{2}}.
\end{equation}
We express $a$ and $n$ in terms of $E_n$ and expand the latter in
powers of $\lambda$ and find
\begin{subequations}
\begin{eqnarray}
E_n^{(0)}&=&(2n+\frac{1}{2})\hbar\omega\label{3.21},\\
E_n^{(1)}&=&-\frac{1}{\sqrt\pi}\sqrt{\frac{m\omega}{\hbar}}
\frac{\Gamma(n+\frac{1}{2})}{\Gamma(n+1)}\label{3.22},\\
E_n^{(2)}&=&-\frac{m}{2\pi^2\hbar^2}
\left[\frac{\Gamma(n+\frac{1}{2})}{\Gamma(n+1)}\right]^2
\left[{\psi{(n+1)}}-{\psi{(n+\frac{1}{2})}}\right],\label{3.23}
\end{eqnarray}
\end{subequations}
where
\begin{subequations}
\begin{align}\label{3.24}
{\psi{(n+1)}}&=\frac{\Gamma'(n+1)}{\Gamma(n+1)},\\
{\psi{(n+\frac{1}{2})}}&=\frac{\Gamma'(n+\frac{1}{2})}{\Gamma(n+\frac{1}{2})}.
\end{align}
\end{subequations}

On the other hand, if we substitute the values of the energy
eigenvalues and eigenfunctions of the unperturbed Hamiltonian in Eq.
(4), we can express $E_n^{(2)}$ as
\begin{equation}\label{3.26}
E_n^{(2)}=\frac{1}{2\pi}\frac{m}{\hbar^2}\sum_{l\neq{n}}\left[\frac{1}{2^{2(n+l)}}
\frac{(2l)!(2n)!}{(l!)^2(n!)^2}\frac{1}{(n-l)}\right].
\end{equation}
From Eqs.~(\ref{3.23}) and (\ref{3.26}), we find
\begin{equation}\label{3.27}
\sum_{l\neq{n}}\left[\frac{1}{2^{2(n+l)}}
\frac{(2l)!(2n)!}{(l!)^2(n!)^2}\frac{1}{(n-l)}\right]=-\frac{1}{\pi^2}
\left[\frac{\Gamma(n+\frac{1}{2})}{\Gamma(n+1)}\right]^2
\left[{\psi{(n+1)}}-{\psi{(n+\frac{1}{2})}}\right],
\end{equation}
which is a new series with a known sum at the right hand side and
may be of mathematical interest.

We end this section with a brief look at the one-dimensional
hydrogen atom. In this case we place the $\delta$-function potential
at $x=a$, so that the Hamiltonian is
\begin{equation}\label{6.5}
H=-\frac{\hbar^2}{2m}\frac{d^2}{dx^2}-\frac{e^2}{|x|}-\lambda\delta(x-a).
\end{equation}
The eigenvalue equation is with $E=-|E|$
\begin{equation}\label{6.6}
\frac{d^2\psi}{d\rho^2}+\frac{2\alpha}{\rho}\psi+\lambda\sqrt{\frac{2m}{\hbar^2|E|}}\delta(\rho-ka)\psi=\psi,
\end{equation}
where $\alpha=e^2 \sqrt{m/2\hbar^2|E_0|}=e^2m/\hbar^2k$. The
wavefunction is continuous at $\rho=ka$ and
\begin{equation}\label{6.7}
\frac{d\psi}{d\rho}\Big|_{+}-\frac{d\psi}{d\rho} \Big
|_-=-\lambda\sqrt{\frac{2m}{\hbar^2|E|}}\psi
\end{equation}
at $\rho=ka$. For $\rho<ka$, $\psi$ is well behaved at the origin
and vanishes at infinity. We have
\begin{equation}\label{6.11}
\psi=
\begin{cases}
A\rho{e}^{-\rho}M(1-\alpha,2,2\rho), & \rho<ka \\
\psi=B\rho{e}^{-\rho}U(1-\alpha,2,2\rho) & \rho>ka.
\end{cases}
\end{equation}
Here
\begin{equation}
\label{6.9} M(a,b,z)=1+\frac{az}{b}+\frac{(a)_2z^2}{(b)_2}+ \ldots,
\end{equation}
and
\begin{subequations}
\begin{align}
U(1-\alpha,2,2\rho)&=\lim_{\epsilon\rightarrow{0}}U(1-\alpha,2+\epsilon,2\rho)\\
&=\lim_{\epsilon\rightarrow{0}}\frac{\pi}{\sin
\pi\epsilon}\left[{\frac{M(1-\alpha,2,2\rho)}
{\Gamma(-\alpha)}}-\frac{1}{2\rho}
\frac{M(-\alpha,-\epsilon,2\rho)}{\Gamma(1-\alpha)\Gamma(-\epsilon)}\right]
\end{align}
\end{subequations}
and $(\beta)_n=\beta(\beta+1)(\beta+2) \cdots (\beta+n+1)$. We find
the eigenvalue equation
\begin{align}\label{6.14}
-M(1-\alpha,2,2ka)U'(1-\alpha,2,2ka)
&+M'(1-\alpha,2,2ka)U(1-\alpha,2,2ka) \nonumber \\
&{}= \frac{m\lambda}{\hbar^2k}M(1-\alpha,2,2ka)U(1-\alpha,2,2ka).
\end{align}

Now we can proceed to find the second-order energy shift. Although
it is not instructive to show more details, outlining the procedure
for the hydrogen atom is useful because it is a prototype for
problems in higher dimension. In such situations, if we want to
perturb the exactly soluble model by a $\delta$-function, it is
necessary to put the one dimensional $\delta$-function at some $r>0$
rather than at $r=0$, where it has no influence.

\section{Discussion}\label{VII}
To obtain a feeling for the convergence properties of the series
that we have found, we present some numerical results. We first
consider Eq.~(\ref{1.17}) and rewrite it in the form
\begin{equation}\label{79}
\sum_{l\neq{n}}\frac{1}{l^2-n^2}= \frac{1}{4n^2}\;\;\quad \mbox{$(l,
n$ odd)}
\end{equation}
Table~\ref{tab1} shows how the sum on the left approaches $1/4n^2$
in two sample cases, e.g. at $n=1$ and $n=5$, with increasing number
of terms. We can generate other series of this sort by taking other
values of $p\,(\neq \frac{1}{2})$ in Eq.~(\ref{1.16}). It is also
interesting to determine what happens to the sum in Eq.~(\ref{79})
if the restrictions on $l$ and $n$ are removed. We can easily find
 that such a series converges to $-1/4n^2$:
\begin{equation}\label{80}
\sum_{l=0(\neq n)}\frac{1}{l^2-n^2}=- \frac{1}{4n^2} \quad (n\neq
0).
\end{equation}
To verify, one may proceed as follows:

\begin{eqnarray}
\sum_{l=0(\neq n)}\frac{1}{l^2-n^2}&=&\lim_{N\rightarrow \infty}
\left(\sum_{l=0}^{n-1}\frac{1}{l^2-n^2}+\sum_{l=n+1}^{N}\frac{1}{l^2-n^2}\right)\nonumber\\
&=&\frac{1}{2n}\sum_{l=0}^{n-1}\left(\frac{1}{l-n}-\frac{1}{l+n}\right)+
\frac{1}{2n}\lim_{N\rightarrow \infty}\sum_{l=n+1}^{N}\left(\frac{1}{l-n}-\frac{1}{l+n}\right)\nonumber\\
&=&-\frac{1}{2n}\sum_{j=1}^{n}\frac{1}{j}-\frac{1}{2n}\sum_{j=n}^{2n-1}\frac{1}{j}
+\frac{1}{2n}\lim_{N\rightarrow
\infty}\left(\sum_{j=1}^{N-n}\frac{1}{j}-
\sum_{j=2n+1}^{N+n}\frac{1}{j}\right)\nonumber\\
&=&-\frac{1}{2n}\frac{1}{n}-\frac{1}{2n}\sum_{j=1}^{2n-1}\frac{1}{j}
+\frac{1}{2n}\lim_{N\rightarrow
\infty}\left(\sum_{j=1}^{N-n}\frac{1}{j}
-\sum_{j=2n+1}^{N+n}\frac{1}{j}\right)\nonumber\\
&=&-\frac{1}{2n^2}+\frac{1}{4n^2}+\frac{1}{2n}\lim_{N\rightarrow
\infty}\left(\sum_{j=1}^{N-n}\frac{1}{j}
-\sum_{j=1}^{N+n}\frac{1}{j}\right)\nonumber\\
&=&-\frac{1}{4n^2},\nonumber
\end{eqnarray}

as, at the limit $N\rightarrow \infty$, $N\pm n\approx N$.
Consequently, the terms within parentheses tend to zero. \\We now
turn our attention to Eq.~(\ref{1.23}) which simplifies to the form
\begin{equation}\label{81}
\sum_{l\neq n}\frac{\sin l p\pi \sin{\frac{l\pi{x}}{L}}}{(l^2-n^2)}
=\frac{1}{4n^2}\sin n p\pi\cos 4np\pi
\sin{\frac{n\pi{x}}{L}}-\frac{\pi x}{2nL} \sin n
p\pi\cos{\frac{n\pi{x}}{L}},
\end{equation}
where $l$ and $n$ are odd integers, $0\leq p\leq 1$ and $0\leq x\leq
pL$. Unlike Eq.~(\ref{1.16}), there is an additional variable $x$ so
that we can find many interesting infinite series with closed form
answers. For example, we consider $p= 1/2$, $x=L/4$, and $n=1$ in
Eq.~(\ref{81}) that yields
\begin{equation}\label{series}
\frac{1}{3^2-1}+\frac{1}{5^2-1}-\frac{1}{7^2-1}-\frac{1}{9^2-1}+\frac{1}{11^2-1}+\frac{1}{13^2-1}+
\ldots =\frac{\pi-2}{8}.
\end{equation}
If we group terms that are adjacent and have the same sign, we
obtain
\begin{equation}\label{series1}
\sum_{k=1}^{\infty}\frac{-1^{k+1}}{4k}\left(\frac{1}{4k-2}+\frac{1}{4k+2}\right)=\frac{\pi-2}{8},
\end{equation}
with no restriction on $k$. This series is interesting because it
shows a sawtooth convergence. Alternate partial sums provide upper
and lower bounds, which become gradually closer. For such series, it
is well known that averaging two adjacent terms leads to better
convergence. Table~\ref{tab3} shows some results for $\pi$ with a
small number of terms in Eq.~(\ref{series1}). We denote the sums by
$S(j)$ and construct $\overline{S(j)}=\frac{[{S(j)+S(j-1)}]}{2}$ to
see how the average sum approaches better the desired value. By
increasing the number of terms in Eq.~(\ref{series1}), we can obtain
very good estimates of $\pi$ as shown in Table~\ref{tab4}. Note that
the series (\ref{series}) or (\ref{series1}) is not commonly used to
estimate $\pi$. The well known Gregory-Leibniz series converges very
slowly \cite{wolfram}. Many other series for $\pi$ are available
with varying convergence properties (see, for example,
Ref.~\onlinecite{wolfram}). Series (67) may be a useful addition to
such a list.

We note finally that the infinite sums in Eqs.~(\ref{1.16}) and
(\ref{series1}) for specific values of the parameters such as $p$
and $n$ or the variable $x$, have a common property: They have the
same asymptotic behavior as the series (\ref{1.17}). Hence, we can
be sure about their convergence, in view of Eq.~(\ref{1.17e}).

\section{Conclusion}\label{VII}
Exactly soluble potentials remain exactly soluble when we add a
$\delta$-function or a set of $\delta$-functions (see Problem~3 in
Sec.~VI for an example). In the three cases we studied, the problem
can be solved in two different ways. An exact solution can be
obtained by joining the piecewise exact results for the wavefunction
at the $\delta$-discontinuities. Or we can use perturbation theory
in the strength of the $\delta$-function to solve the problem to any
desired order. Because the second-order perturbation theory result
is usually in the form of an infinite series, we can use the exact
solutions to determine the sums of such series, and exhibit their
convergence properties. We can consider other exactly soluble
potentials as well. \cite{Pronchik} New infinite sums with known
closed form answers may be found. Recently, an alternative scheme
has been put forward using Green's function
techniques.\cite{Sukumar}

By choosing appropriate examples, we can sometimes get insight into
physics of certain problems that are not usually considered. The
role of the continuum in the second order shift for the finite well
is a case in point. In Problem~1 we explore the effect of a single
attractive $\delta$-function. We have seen that below a critical
length, the bound state of the $\delta$-function disappears. Two
attractive $\delta$-functions can be used to set up a toy model for
the hydrogen molecular ion. Placing walls in this case on the two
sides allows us to study the effect of confinement on the binding of
a molecule. We can imagine carrying this further and placing a
periodic array of $\delta$-functions interrupted on each side by a
wall. The resulting violation of Bloch's theorem would make an
interesting study.

\section{Suggested Problems}

\textit{Problem 1}. In Sec.~II we considered only the positive
eigenenergies for the $\delta$-function potential in a box. The
attractive $\delta$-function by itself will have a negative energy
bound state. In this problem we explore what happens when the
$\delta$-function is placed in a box. Consider a $\delta$-function
at the center of a box of width $2L$. Repeat the calculation of
Sec.~II, but for a negative energy bound state and show that for a
$\delta$-function of strength $\lambda$, the negative energy bound
state will disappear if $L$ is smaller than $L_c= \hbar^2/m
\lambda$.

\textit{Problem 2}. Consider the perturbation of a particle confined
in a one-dimensional box of length $L$ by an attractive
$\delta$-function potentials at $x=pL$ with $0<p<1$. The
wavefunction is given as
\begin{equation}
\psi(x)=
\begin{cases}
A\sin kx & (0\leq x \leq pL) \\
A\frac{\sin kpL}{\sin k(1-p)L}\sin k(L-x) & (pL\leq x \leq L),
\end{cases}
\end{equation}
where $A$ is a normalization constant. Explore the wavefunction in
the region $pL\leq x \leq L$ to obtain the first order correction to
wavefunction. Also derive the same from perturbation theory and show
that this leads to an interesting sum rule. Is the series
convergent?

\textit{Problem 3}. Consider a particle confined to a
one-dimensional box of length $L$ perturbed by two attractive
$\delta$-function potentials at $x=pL$ and $x=qL$ with $0<p<1$ and
$0<q<1$. (a) Find the second order energy shift and show that it is
always negative when $p=\frac{1}{4}$ and $q=\frac{3}{4}$. (b) Do the
same using Rayleigh-Schr\"{o}dinger perturbation theory and compare
the result with the above. Does this series converge? (c) Find the
first order correction to the wavefunction and show how this
correction gives three important sum rules which are convergent.

\textit{Problem 4}. Consider a simple harmonic oscillator and
perturb it by an attractive delta potential at the origin. Use
perturbation theory to find the first order correction to the
wavefunction and show that this correction leads to another sum
rule.

\begin{acknowledgements} We are thankful to an anonymous referee for many constructive suggestions. NB would like to thank the Council of
Scientific and Industrial Research, India for a research fellowship.
\end{acknowledgements}

\newpage

\section*{Tables}

\begin{center}
\begin{table}[h]
\begin{tabular}{|c|c|c|}
\hline
No. of terms & Results for $n=1$ & Results for $n=5$\\
\hline 10 & 0.2273 & -0.013 \\\hline 50 & 0.2451 & 0.0051 \\\hline
100 & 0.2475 & 0.0075
\\\hline 1000 & 0.2498 & 0.00975 \\\hline 10 000 & 0.249975 & 0.009975 \\\hline
100 000 & 0.2499975 & 0.0099975 \\
\hline
\end{tabular}
\caption{\label{tab1}Convergence study of Eq. (63) in two sample
cases }
\end{table}
\end{center}

\begin{center}
\begin{table}
\begin{tabular}{|c|c|c|}
\hline
No. of terms & Results for $\pi$ & Average estimate \\
$[j]$ & $[S(j)]$ & $\overline{[S(j)]}$ \\\hline 1 & 3.33 & \\\hline
2 & 3.07 & 3.20 \\\hline 3 & 3.18 & \\\hline 4 & 3.12 & 3.149
\\\hline 5 & 3.16 & \\\hline 6& 3.13 & 3.144 \\\hline 7 & 3.15 &
\\\hline 8 & 3.13 & 3.143 \\\hline 9 & 3.147 & \\\hline 10 & 3.137 &
3.142 \\\hline 19 & 3.1429 & \\\hline 20 & 3.1404 & 3.1417 \\\hline
\end{tabular}
\caption{\label{tab3}Convergence study of Eq. (67) for estimates of
$\pi$}
\end{table}
\end{center}

\begin{center}
\begin{table}[h]
\begin{tabular}{|c|c|}
\hline $j$&$\overline{S(j)}$ \\\hline 10 & 3.142 \\\hline 100 &
3.141 593 \\\hline 1000 & 3.141 592 654 \\\hline 10\,000 & 3.141 592
653 59 \\\hline
100\,000 & 3.141 592 653 589 793 8 \\
\hline
\end{tabular}
\caption{\label{tab4} Gradually improved estimates of $\pi$ from Eq.
(67)}
\end{table}
\end{center}

\end{document}